\documentclass[twocolumn,pre,showpacs]{revtex4-1}
\usepackage{graphicx}

\begin{document}
\title{Fluctuations of isolated and confined surface 
   steps of monoatomic height}
\author{Walter Selke}
\affiliation{Institut f\"ur Theoretische Physik and JARA-HPC, RWTH Aachen
  University, 52056 Aachen, Germany}

\begin{abstract}
The temporal evolution of equilibrium fluctuations for surface
steps of monoatomic height
is analyzed studying one-dimensional
solid--on--solid models. Using Monte Carlo simulations, fluctuations
due to periphery--diffusion (PD) as well as due to
evaporation--condensation (EC) are considered, both
for isolated steps and steps confined by the presence of
straight steps. For isolated steps, the dependence of the characteristic
power--laws, their exponents and prefactors, on temperature, slope, and
curvature is elucidated, with the main emphasis on PD, taking into
account finite--size effects. The entropic repulsion due to
a second straight step may lead, among others, to an interesting
transient power--law like growth of the fluctuations, for PD. Findings 
are compared to results of previous Monte Carlo simulations and
predictions based, mostly, on scaling arguments and Langevin theory.
\end{abstract}


\maketitle

\section{Introduction}

Fluctuations of steps of monoatomic height
on crystal surfaces have been studied quite extensively in the past nearly
two decades, both experimentally
and theoretically \cite{rev1,rev2,rev3}.

The step fluctuations are described by the equilibrium time correlation
function $G(t) \propto <(h(i,t+t_0)-h(i,t_0))^2>$, averaging over
all step sites, with 
$h(i,t)$ denoting the step position at site $i$ and time $t$. For
\textit{isolated} indefinitely long steps, one finds a power law for the growth
of the fluctuations, $G\propto t^{x}$. The characteristic exponent 
$x$ is observed to depend on the atomic mechanism driving
the step dynamics, for instance,  
periphery diffusion (PD), with $x=1/4$, or evaporation--condensation (EC),with
$x=1/2$. The theoretical approaches include 
Langevin theory, scaling arguments, and
Monte Carlo simulations. On the experimental
side, especially, scanning tunneling microscopy turned out to be a very
effective tool in measuring the temporal evolution of step fluctuations.

For \textit{confined} steps, fluctuations are affected by the presence
of neighboring steps. Fairly recently, Ferrari et al. \cite{fps} 
analyzed the dynamics of the border ledge 
of a crystal facet. Motivated by this analysis, new characteristic 
fluctuation laws have been suggested, in particular, $x= 2/11$ 
for periphery diffusion,
and $x=2/5$ for evaporation--condensation \cite{deg1}. In subsequent
theoretical and experimental
studies \cite{deg2,deg3} periphery diffusion
has been investigated. For instance, in Monte Carlo
simulations for PD, the fluctuating step has been mimiced, in a
toy model, by the one--dimensional
solid--on--solid (SOS) model, with the neighboring step having been assumed to 
be straight. The two steps interact through an effective, long--range
interaction, the entropic
repulsion \cite{gm,fifi}. Quite good agreement between
results of the theoretical
and experimental approaches has
been reported \cite{deg2,deg3,ep}.

The aims of this article, presenting results of extensive
Monte Carlo simulations on one--dimensional SOS models, are as
follows: (1) In case of isolated steps with PD, effects of slope
and curvature of the steps as well as temperature on the 
characteristic exponent $x$ and on the prefactor of the
corresponding power law for $G(t)$ are investigated. Finite-size
and finite-time dependences are closely monitored, which had not
been studied systematically in previous
simulations. In addition, EC dynamics is simulated. (2) In case
of confined steps, the toy model will be reanalyzed for 
step fluctuations driven by PD. The previous Monte
Carlo findings will be scrutinized. We also shall discuss 
briefly EC kinetics for
the toy model and a modification, which
have not been treated before in simulations.

The outline of the article reflects these aims: Following the introduction, 
we shall describe the SOS models and Monte Carlo methods. Thereafter, the
main results will be presented, first
for isolated and, then, for confined steps. The summary will conclude
the article.  

\section{Models and Monte Carlo methods}

A step of monatomic height on a crystal surface with the kink
excitation energy $J$ may be described by a one--dimensional
solid--on--solid (or terrace--step--kink) model, with the Hamiltonian

\begin{equation}
{\cal H}=  J \sum_{i} \vert h(i) - h(i+1) \vert
\end{equation}

\noindent
where the sum runs over all step sites $i$, with the positions of the
step atoms, $h(i)$, taking integer values. In the simulations, steps
of finite length, $L$, are considered. Unless stated otherwise, we employed
pinned boundary conditions, keeping $h(1)$ and $h(L)$ constant
for all times $t$, measured in Monte Carlo steps per site, MCS/S \cite{LanBin}.

For \textit{isolated} steps, three different initial step 
configurations, $h(i,0)= h(i,t=0)$,
have been studied: (a) Flat steps, with 

\begin{equation}
h(i,0)= 0, i=1,...L
\end{equation}

\noindent
(b) Tilted steps, with 

\begin{equation}
h(i,0)= nint(s(i-1)), i=1,...L
\end{equation}

\noindent
where $s$ is the slope of the
step, pinning the steps at $h(1,t)= 0$ and
$h(L,t)= nint(s(L-1)$ for all times $t$. (c) (Circular) curved steps, with

\begin{equation} 
h(i,0)= nint (\sqrt{(L-1)^2/4-((L+1)/2-i)^2}, i=1,..L, 
\end{equation}

\noindent
for steps pinned at $h(1,t)= h(L,t)= 0$, see Fig. 1.

For \textit{confined} steps, fluctuations of the initially flat step, see case
(a), are restricted by the bordering second, non-fluctuating step
at distance $d$. Thence, there is the constraint $h(i,t)<d$ for
all sites and times.  

The step fluctuations may be measured by the equilibrium correlation
function

\begin{equation}
G(t)= 1/L_a \sum\limits_{i} \langle (h(i,t+t_0)- h(i,t_0))^2 \rangle
\end{equation}

\noindent
with $L_a$ being the number of active sites, i.e. $L_a=L-2$ for
pinned boundary conditions (and $L_a=L$ for periodic boundary conditions).
The brackets denote the thermal average. $t_0$ is a time, at
which equilibrium has been reached.

In the Monte Carlo simulations, we study periphery diffusion and
evaporation--condensation. For PD, step atoms are allowed to
move to neighboring sites, for instance, $h(i) \longrightarrow h(i)-1$,
while $h(i+1) \longrightarrow h(i+1) +1$. In contrast to the Kawaski
dynamics, corresponding to PD, EC is realized by Glauber kinetics, where
at randomly chosen site $i$ one tries, randomly, to change the position of
the step, $h(i)$, by $-1$ (detachment or evaporation) or by $+1$ (attachment
or condensation). As usual, the acceptance rate of the various
moves is given by the appropriate Boltzmann factor \cite{LanBin}.
To study finite--size effects, steps with between about
20 and about 400 sites were simulated. To equilibrate the steps, 
at least the first $10^5$ Monte Carlo steps per site were discarded.
Typically, to obtain thermal averages, a few thousand independent
equilibrium configurations were evaluated, with, for each
configuration, a similar number of realizations, to compute 
the time evolution. In this way, one
gets simulation data of
high accuracy. Indeed, errors bars are smaller than symbol sizes
in the figures, and we refrained from displaying them.

\section{Isolated steps}

We first present results of Monte Carlo simulations on the time
evolution of the equilibrium step fluctuations, $G(t)$, for
\textit{isolated} (a) flat, (b) tilted, and (c) curved
steps, applying periphery diffusion
for all types, and evaporation--condensation for flat and
tilted steps. The
steps are pinned at the ends, i.e. $h(1,t)$, and $h(L,t)$ are
fixed, to avoid, for EC, effects due the motion of the
entire step \cite{bs}. Note that
for curved steps, the equilibrium shape is not circular, as
depicted in Fig.1.

\begin{figure}
\resizebox{0.95\columnwidth}{!}{%
  \includegraphics{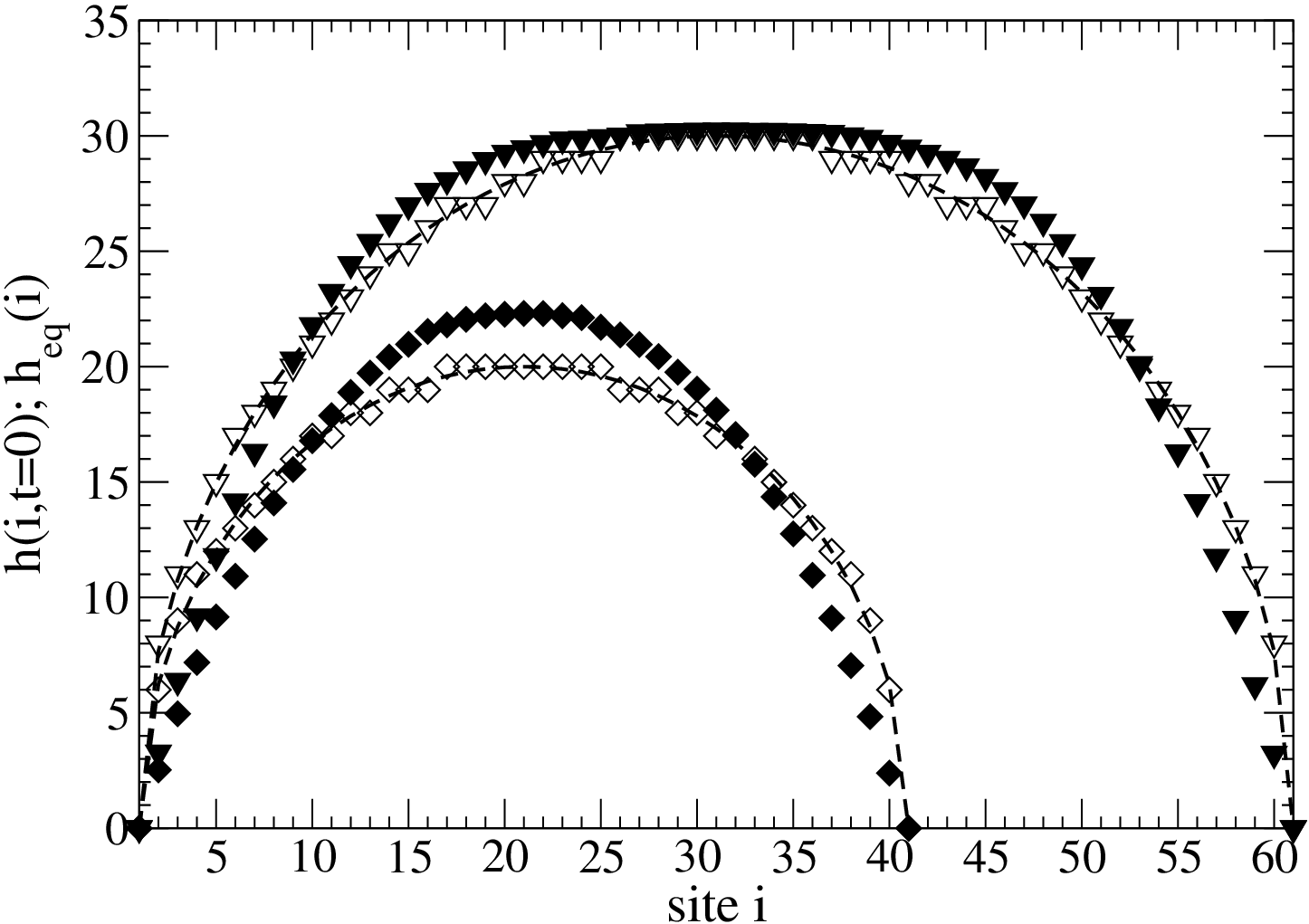}
}
\caption{Initial, $h(i,t=0)$, (open symbols) and PD
 equilibrium, $h_{eq}(i)$, (full symbols) positions of
 curved steps at $k_BT/J= 3.0$ for $L$= 41 and 61.}
\label{fig:1}
\end{figure}

We are interested in the value of the characteristic
exponent $x$ in the expected power--law for the growth of
$G(t) \propto t^x$. Using Langevin theory
and simulations, $x$
has been found to be 1/4 for fluctuations driven by PD and 1/2
in case of EC \cite{rev1,rev2,bs,bladux,eins1,sza}. The latter result follows
from exact arguments as well \cite{au}.

To determine $x$ from simulation data, one may monitor
the effective exponent $x_{eff}$, in its discretized form \cite{ps}

\begin{equation}
x_{eff}(t,L)= \ln (G(t_j)/G(t_{j-1}))/ \ln (t_j/t_{j-1})
\end{equation}

\noindent
with $t= \sqrt{t_j t_{j-1}}$, where $t_j$ refers to the discrete time
scale, measured in MCS/S.

Indeed, after a diffusive--like behavior, $x_{eff} \approx 1$, at very short
times \cite{bs}, the effective exponent is observed to decrease
monotonically rather rapidly to about 1/4 for PD, and to about 1/2 for EC, 
for flat, tilted, and curved steps, as illustrated in
Fig. 2. Note that $x_{eff}$ continues to
decrease monotonically, approaching 
eventually zero, due to the fact that $G(t)$ saturates for pinned
steps of finite length. Let us denote the time needed to pass
through the characteristic value, $c=$ 1/4 or 1/2, by $t_{c}$, i.e.
$x_{eff}(t_{c})= c$. That time
depends on temperature, slope, curvature, and, perhaps, most 
interestingly, the length of the step. Actually, we find strong evidence
that $t_c$, for sufficiently long steps, increases with $L_a= L-2$
in form of a power--law, $L_a^{\gamma}$, with
$\gamma$ being roughly 1.5 for EC and being roughly 0.5 for
EC. Steps with up to about 400 sites have been
studied. The observed finite--size behavior suggests that
$G(t)$ satisfies the simple power--law, $G(t) \propto t^x$,
asymptotically in time in the thermodynamic
limit, $L_a \longrightarrow \infty$. 

\begin{figure}
\resizebox{0.90\columnwidth}{!}{%
  \includegraphics{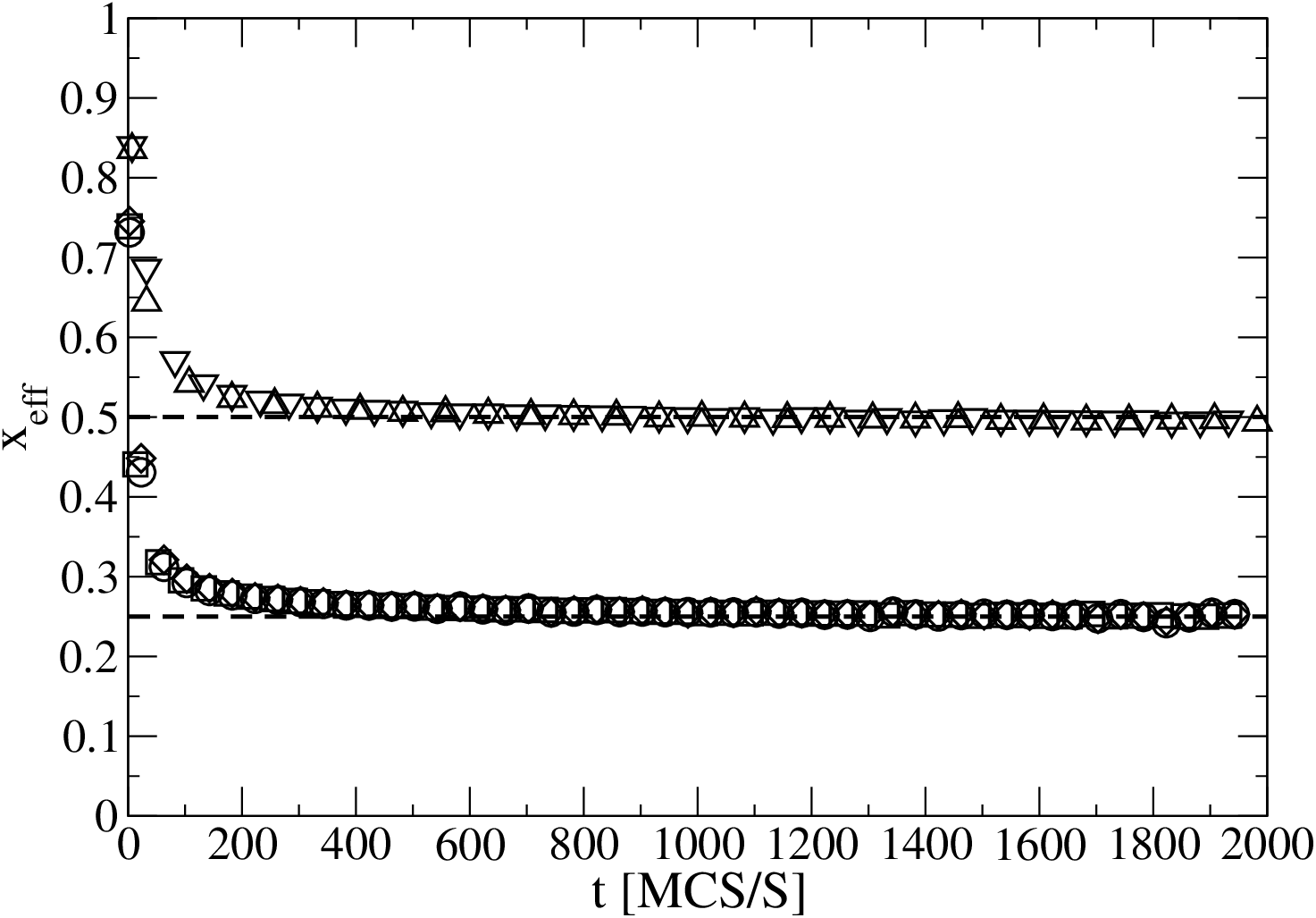}
}
  \caption{Effective exponent $x_{eff}$ at $k_BT/J =3.0$ for PD, for
   flat (circle), tilted, with slope $s=1$, (square) and curved (diamond)
   steps, $L=101$, as well as for EC, for
   flat (triangle up) and tilted, with $s=1$, (triangle down) steps, $L=201$.}
\label{fig:2}
\end{figure}

The prefactor, $a$, in front of the asymptotic power--law 
for $G$, i.e. $G(t)=at^x$, turns out to depend significantly on
temperature and the type of the step,(a)-(c). To our knowledge, $a$ had
not been analyzed in previous simulations of
step fluctuations. To take into account finite--size effects, we
consider the effective prefactor $a_{eff}$

\begin{equation}
a_{eff}(t,L_a)= G(t)/t^x
\end{equation}

\noindent
with $x$= 1/4 for PD and 1/2 for EC. $L_a$ is the number of
active sites. The effective prefactor is found to have its, in
general plateau--like, maximum, $a_p(L_a)$, at the time $t_c$, at
which the corresponding effective exponent $x_{eff}$ passes
through $c$= 1/4, for PD, or 1/2, for EC. The prefactor
$a$ = $a_p(\infty)$ is then reached for $L_a,t \longrightarrow \infty$.
\begin{figure}
\vspace{1.0cm}
\resizebox{0.90\columnwidth}{!}{%
  \includegraphics{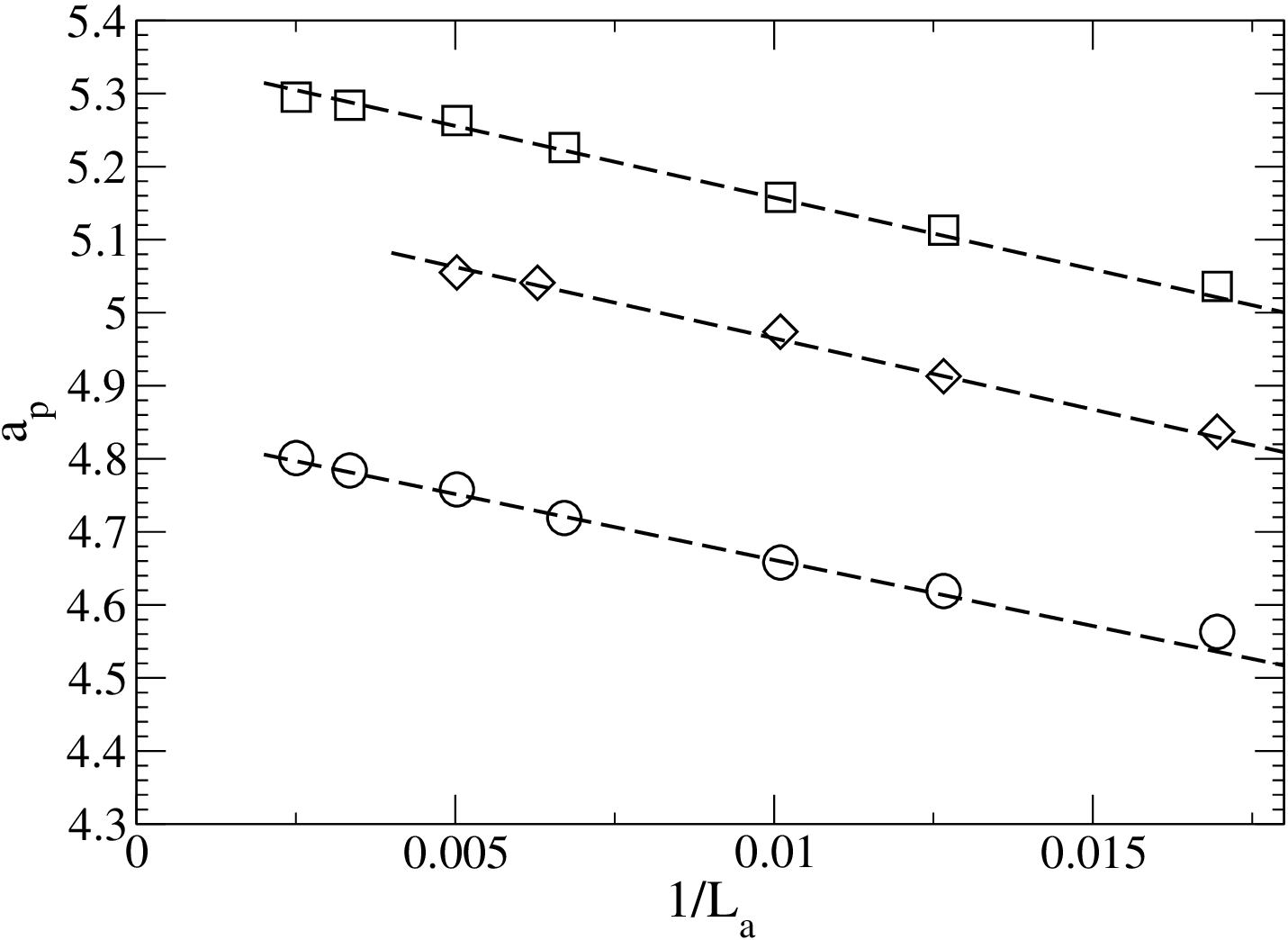}
}
\caption{Effective prefactor $a_p$ vs. inverse number of active
  sites $1/L_a$ for PD for flat (circle), tilted, with slope $s=1$, (diamond)
  and curved (square) steps at $k_BT/J= 3.0$.}
\label{fig:3}
\end{figure}

$a_p(L_a)$ displays an interesting and simple finite--size behavior, as
illustrated, for PD, in Fig. 3. For sufficiently long steps, $a_p(L_a)$ is
found to grow, for all types of steps, (a)-(c), like 
$a_p(L_a)= a_p(\infty)- d/L_a$, $d$ depending on temperature 
and on the type of step. Thence, one
may easily estimate the prefactor $a(T)$ from the simulation data.

Results of such extrapolations for initially flat steps
at various temperatures are shown in Fig. 4. We compare
our Monte Carlo findings to predictions of the Langevin
theory \cite{bew1},

\begin{equation}
G(t)= 0.46... ((k_BT)^3 \Gamma_h/\tilde \beta^3))^{1/4}t^{1/4}
\end{equation}

\noindent
with the step stiffness $\tilde \beta$ being

\begin{equation}
\tilde \beta= 2k_BT \sinh^2(J/2k_BT)
\end{equation}

\noindent
The step mobility $\Gamma_h$ follows, apart
from a proportionality factor, from the fraction of
successful Monte Carlo attempts. In Fig. 4, we
fixed this factor by equating the Langevin predictions and
the Monte Carlo findings at $k_BT/J =1$. Obviously, we observe
a very good agrement for the non--trivial temperature
dependence of the simulated prefactor $a$
with the prediction based on Langevin theory.

\begin{figure}
\vspace{1.5cm}
\resizebox{0.90\columnwidth}{!}{%
  \includegraphics{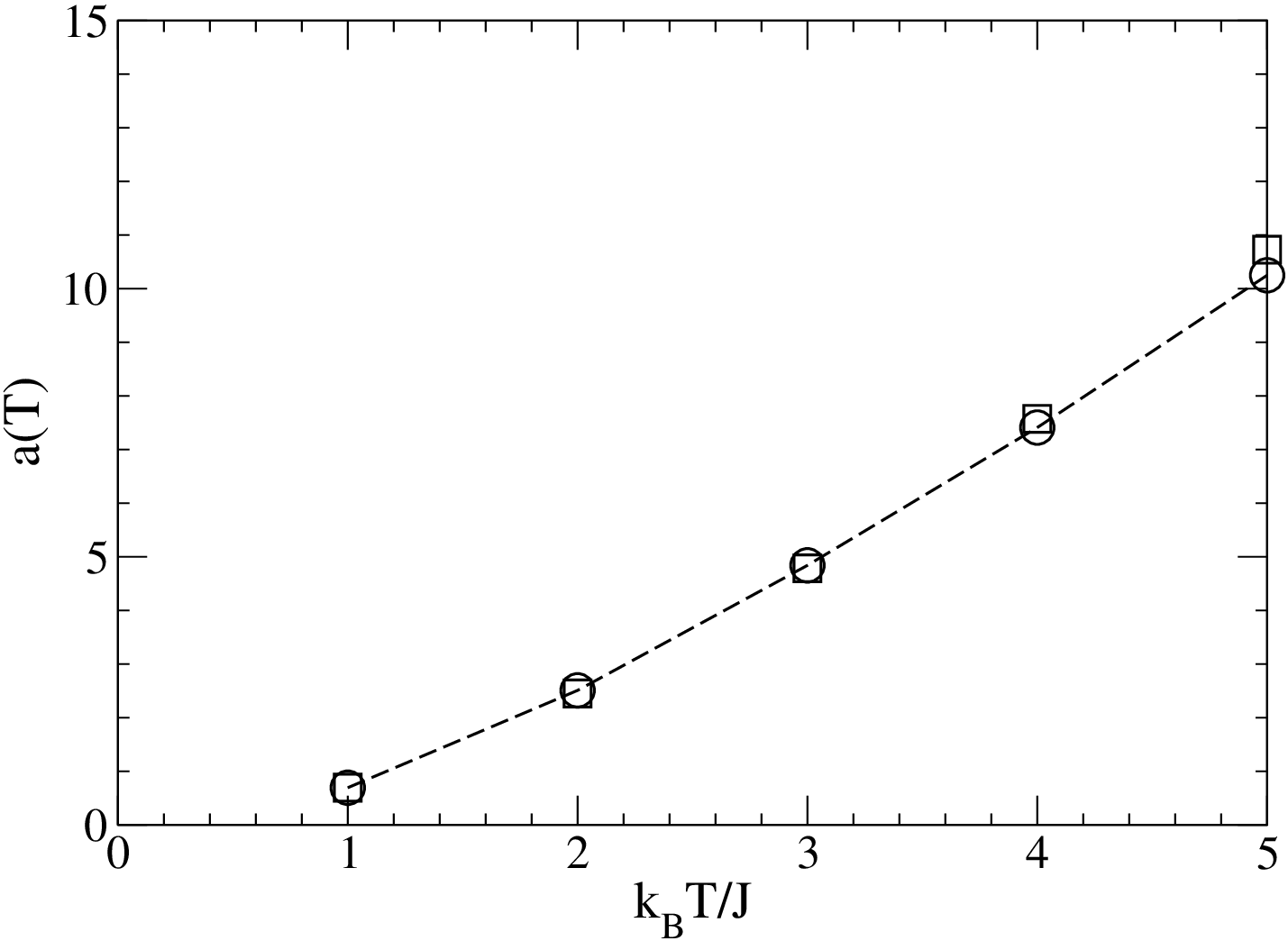}
}
  \caption{Prefactor $a$ of the step fluctuations for PD, at various
 temperatures $k_BT/J$, comparing Monte Carlo findings (circle), extrapolated
 to the thermodynamic limit, with predictions based on Langevin
 theory (square).}
\label{fig:4}
\end{figure}

It seems worthwhile to note
that the prefactor $a$ increases monotonically not only with the temperature
but also with the slope of the steps, $a(s)$, see
Fig. 5.  Consequently, it
is significantly larger for curved than for flat steps. Actually, $a$
is observed to depend essentially
on the energy of the step. For instance, for tilted steps, we obtained

\begin{equation}
a(s)/a(s=0) \approx E(s)/E(s=0)
\end{equation}

\noindent
where $E(s)$ is the thermal equilibrium energy of steps with the slope $s$.

Because of their pinning at the ends, the steps tend to fluctuate more
strongly in the center, as we observed when monitoring the
local fluctuations. Of course, this boundary effect is expected
to play no role in the limit of indefinitely long steps.

\section{Confined steps}

We now consider fluctuating, initially flat steps with periphery
diffusion, described
by the one--dimensional SOS model, Eq.(1), confined by a second
step. The confining step is located at distance $d$ from
the initial step position $h(i,t=0) =0$, i.e. for all times one has
$h(i,t) <d$. A typical Monte Carlo equilibrium configuration
for $d=4$ and periphery diffusion is depicted in Fig. 6.

\begin{figure}
\resizebox{0.90\columnwidth}{!}{%
  \includegraphics{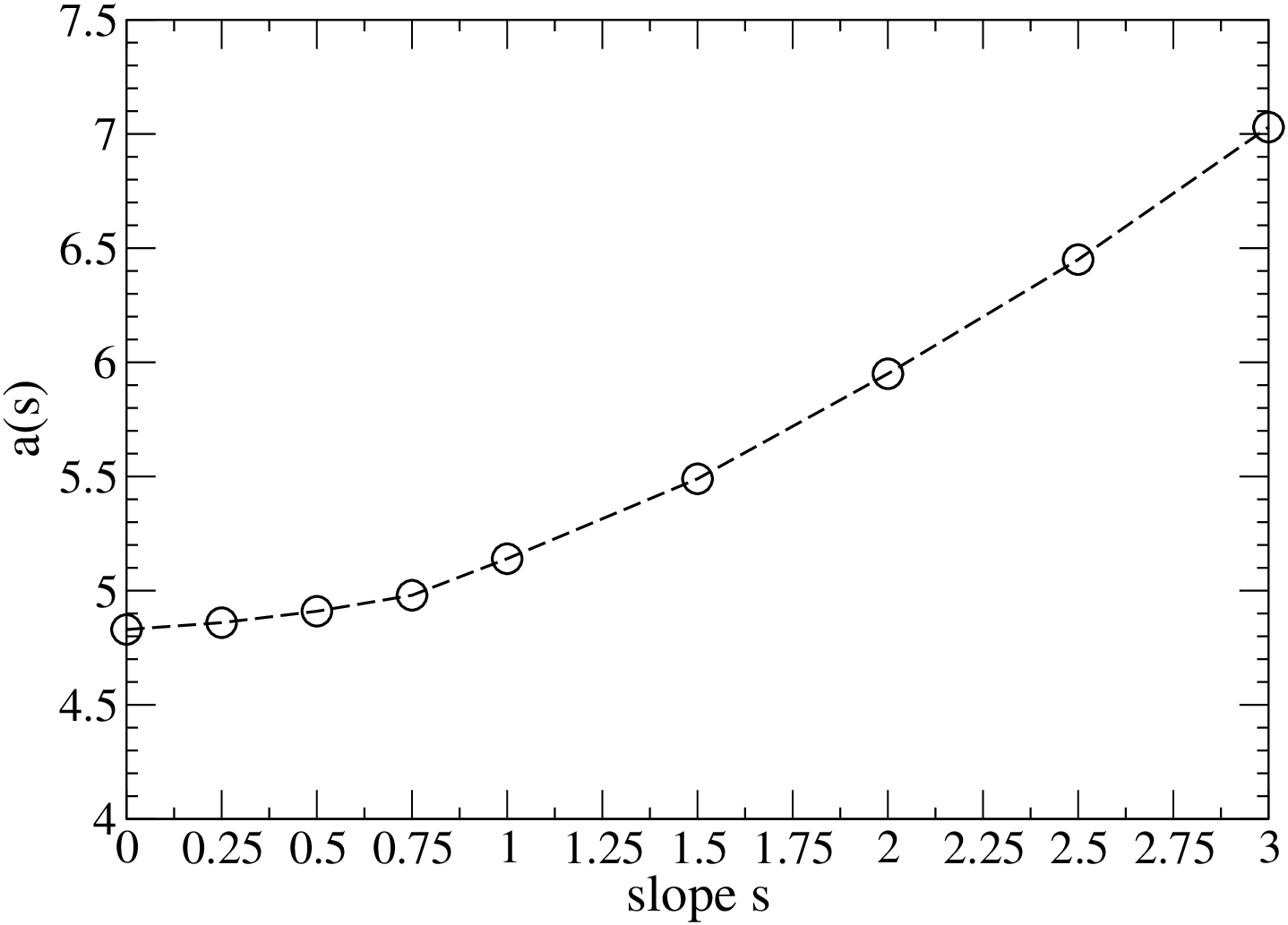}
}
\caption{Slope, $s$, dependence of the (extrapolated) prefactor
 $a$ for tilted steps with PD at 
 $k_BT/J= 3.0$. }
\label{fig:5}
\end{figure}
\begin{figure}
\vspace{0.8cm}
\resizebox{0.90\columnwidth}{!}{%
  \includegraphics{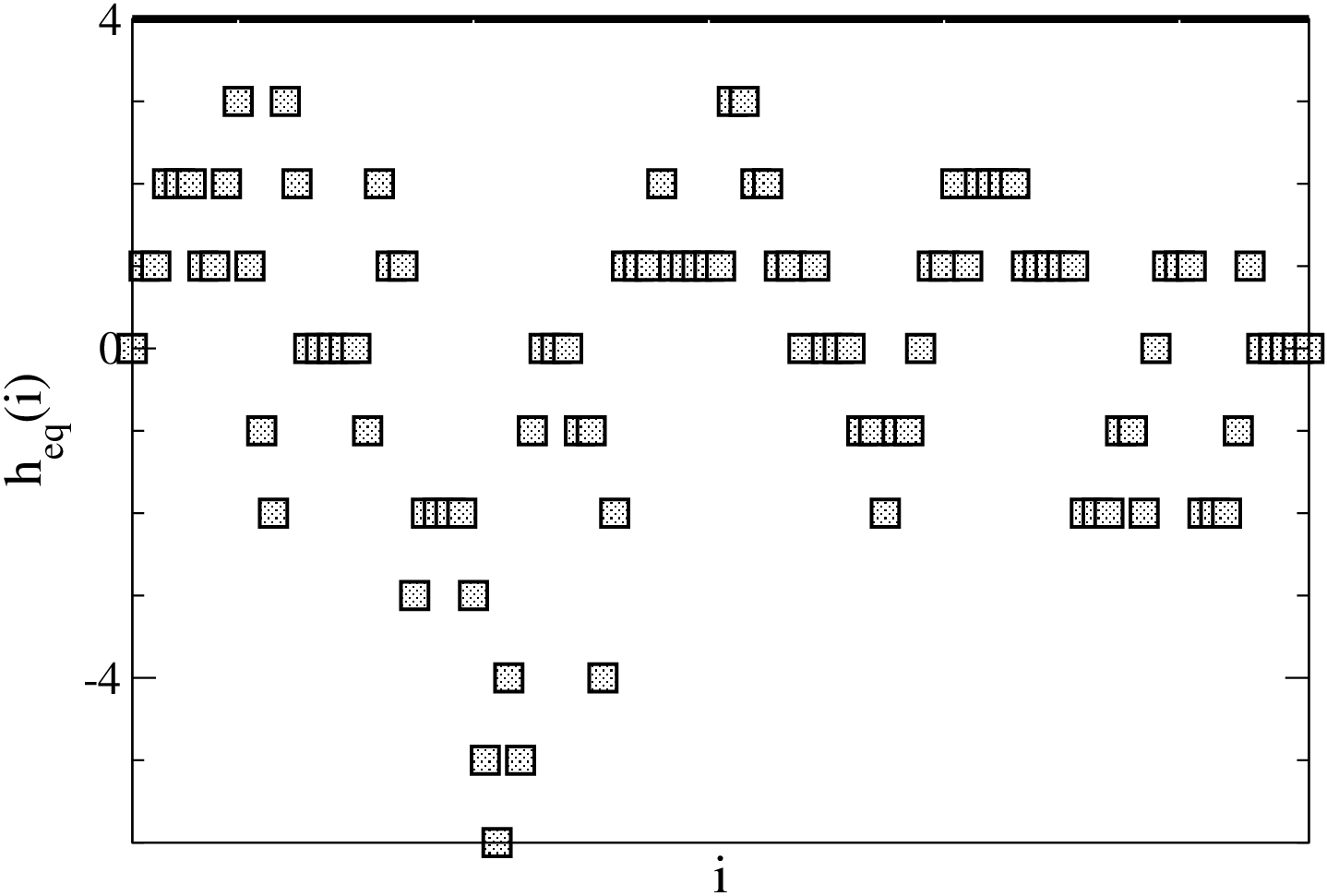}
}
\caption{Typical Monte Carlo equilibrium configuration, PD, for a confined
 step with $d=4$ at $k_BT/J= 1.0$. }
\label{fig:6}
\end{figure}

In previous Monte Carlo studies of this toy model \cite{deg2,deg3}, periodic
boundary conditions have been employed, considering PD
at fixed temperature, $k_BT/J=1.0$, and step length $L=100$. Depending
on the separation distance $d$, three different regimes have
been identified. For small $d$, $d \le 3$, the step fluctuations, $G(t)$,
have been suggested to grow logarithmically with time. For
sufficiently large distances, $d \ge12$, $G(t)$ is found to
behave as for isolated steps, where its power--law like increase is
characterised by the exponent $x= 1/4$. In the intermediate range
of $d$, power--law dependences are observed, with the
exponent $x$ being close to 2/11, as may follow from
theoretical descriptions for the dynamics of the border ledge 
of a crystal facet \cite{fps,deg1}. However, deviations from the
simple power--law have been found to occur, possibly, due to,
for instance, crossover effects related to the separation distance
$d$ \cite{deg2,deg3,ep}. No simulations had been done for
EC, where scaling arguments and Langevin theory \cite{fps,deg1}
may suggest $x$= 2/5, instead of 1/2 as for isolated steps.

The aim of the present Monte Carlo study is to study the toy model
in more detail, investigating, among others, 
finite--size effects, the deviations from simple-power laws and the
long--time behavior of $G(t)$. Mostly, we employed pinned boundary
conditions.

\begin{figure}
\vspace{0.9cm}
\resizebox{0.90\columnwidth}{!}{%
  \includegraphics{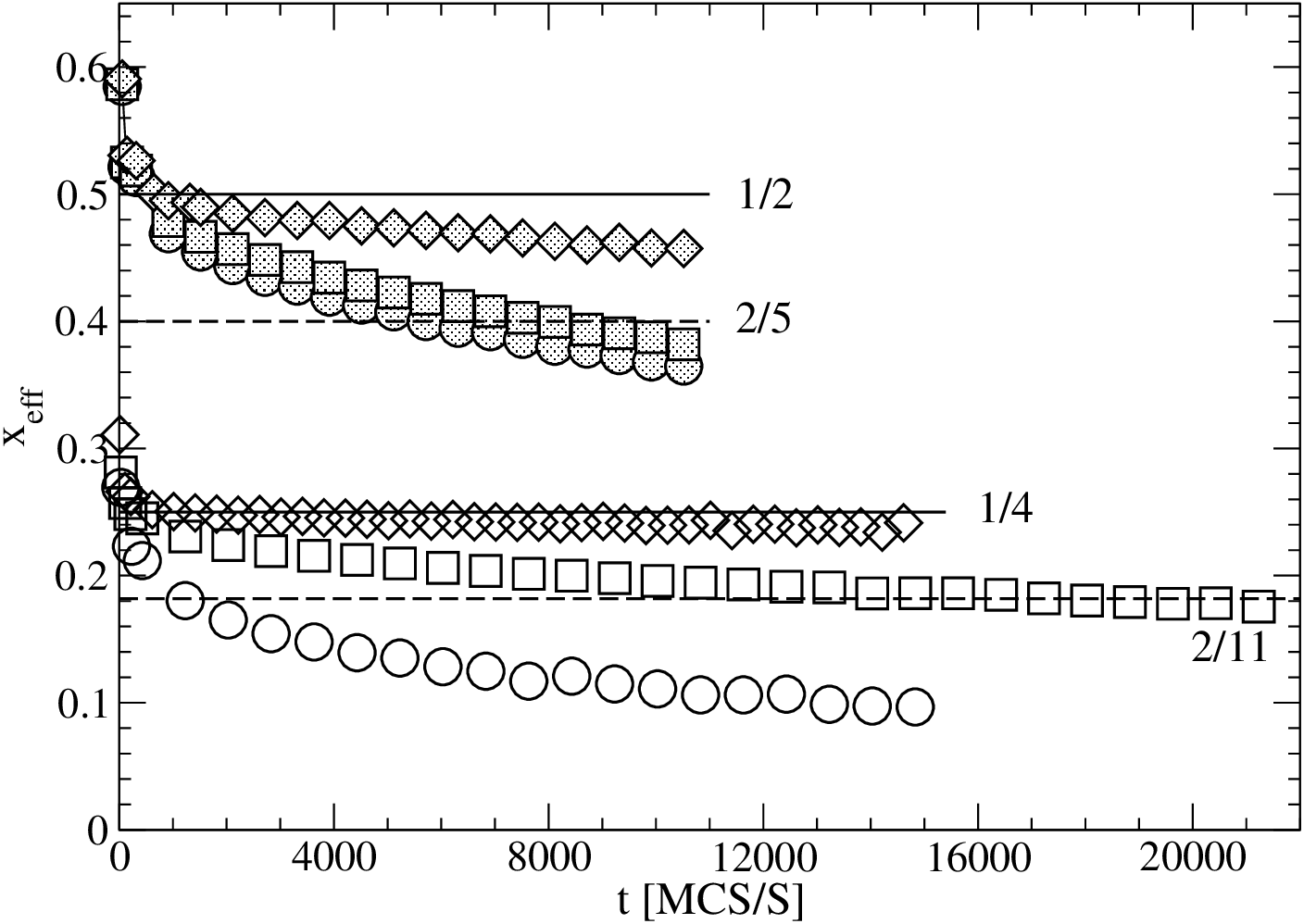}
}
\caption{Effective exponent $x_{eff}$ for confined steps of length
   $L=201$, with PD (open symbols) at $k_BT/J= 1.0$, for
   $d$= 3 (circle), 5 (square),
   and 16 (diamond), and with EC (shaded symbols) at $k_BT/J= 3.0$, for
   $d=3$ (circle), 20 (square), and 300 (diamond).}
\label{fig:7}
\end{figure}
\begin{figure}
\vspace{0.8cm}
\resizebox{0.90\columnwidth}{!}{%
  \includegraphics{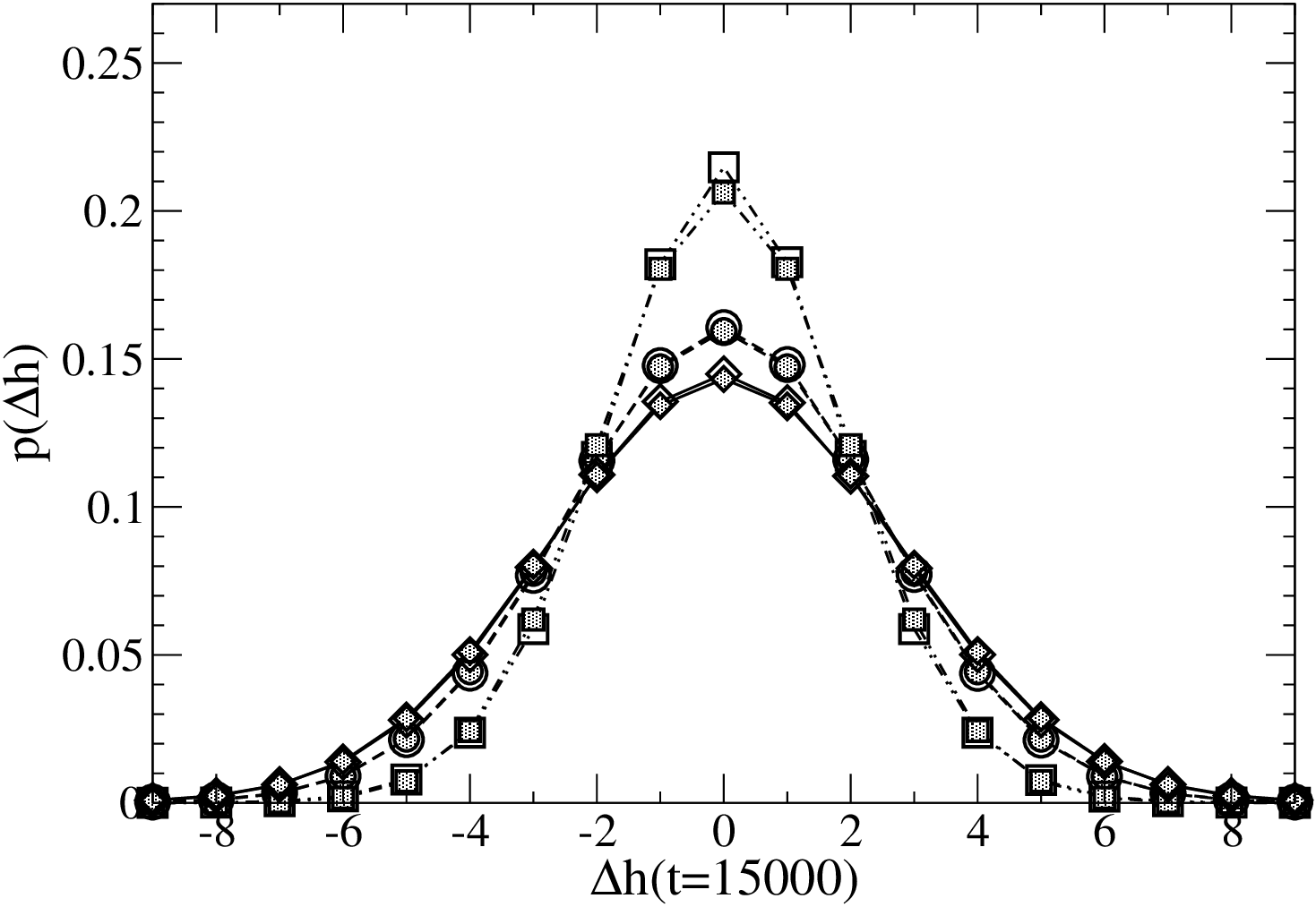}
}
\caption{Histograms of $(\Delta h)$  for pinned isolated (diamond) and confined
 steps with $d=3$ (square) and 5 (circle) of length $L=201$ at
 $k_BT/J= 1.0$, using PD, comparing simulation data (open symbols) at
 $t=15000$ MCS/S to
 Gaussian distributions (shaded symbols) with the standard deviation set equal
 to the fluctuation length.}
\label{fig:8}
\end{figure}

Let us first deal with PD. In particular, we monitored the dependence
of the effective 
exponent $x_{eff}(t)$ on step length, temperature, and
separation distance. At first sight, one may
distinguish three regimes, (i)-(iii), see below, following and
generalizing the
previous analysis \cite{deg2}. Examples
are depicted in Fig. 7.

In all cases, at very early times, $G(t)$ exhibits, as for isolated
steps, a diffusive--like behavior
due to independent moves at randomly chosen sites, with
$x_{eff}$ near one. A much finer time resolution than the
one shown in Fig. 7, would be needed to demonstrate that behavior. Because
of the step stiffness, the growth of the 
fluctuations then slows down \cite{bs}, with $x_{eff}$ decreasing more
gradually and, always, monotonically. Now, (i) the
effective exponent may approach rather quickly zero, without
any striking anomalies. This may happen when
the steps are rather short, independent of $d$, because $G(t)$
saturates quite soon. This rather trivial effect is
not displayed in Fig. 7. It may also also happen for
longer steps and fairly small $d$, as depicted in Fig. 7, possibly, due to
the suggested logarithmic rise of $G(t)$ \cite{deg2,deg3}. On the
other hand, (ii) for sufficiently large values
of $d$, $G(t)$ seems to grow
as for isolated steps, with $x_{eff}$ being, in quite extended time
intervals, close to 1/4. The
most interesting case is
encountered for the intermediate regime (iii), where, again
for quite extended time intervals, one observes a power--law like
behavior, with $x_{eff}$ being close to 2/11. Certainly, for a
more quantitative description, temperature
will play an important role as well, as will be discussed below.

Clearly, in general, $G(t)$ is 
expected to be, asymptotically, finite, in all three regimes of
the toy model, with
the corresponding fluctuation length  $g(t)= \sqrt{G(t)}$ being
bounded by the distance $d$. In any event, it is of much 
interest to clarify the conditions, under which one may observe
typical features of the three regimes (i)-(iii).

We checked, in regime (i), whether there is evidence of the 
logarithmic growth of $G(t)$, which had been suggested
to occur, as in a step train, for small $d$ \cite{rev1,deg2,deg3}.   
We assumed the form
$G(t)= b\ln (t)$ and monitored
the effective prefactor $b=b(t)$. In fact, for instance, for $d=3$ at
$k_BT/J$= 1.0 and $L=201$, see Fig. 7, the prefactor is found
to be nearly constant in a quite large time
range, measured in MCS/S, $5000 \lesssim t \lesssim 25000$. Eventually,
it starts to decrease, due to finite--size saturation of the
fluctuations.

The time window increases with the length of the steps. We tend
to conclude that the fluctuations satisfy, over a fairly large
time interval, the logarithmic growth law. We note in passing that
there is no indication of logarithmic behavior for small steps, 
where $x_{eff}$ goes quickly to zero caused by the fast saturation
of $G(t)$. At higher temperatures, regime(i) extends to a
larger range of separation distances $d$, as will be explained below.

Obviously, in regime (ii), where $x_{eff} \approx 1/4$, the
fluctuation length of isolated steps has to be small compared to
the separation distance $d$. For sufficiently long
steps, it may happen in large time windows, see Fig. 7, but 
asymptotically, $L,t \longrightarrow \infty$, the growth of $G(t)$ 
will be ultimately limited by $d$.

In the, perhaps, most interesting regime of moderate distances
$d$, regime (iii), we observe, again in intermediate time intervals, a
power--law like behavior of $G(t)$, with exponent close to
2/11 for PD, as illustrated in Fig. 7. The value has
been argued to reflect the
long--range interactions of the fluctuating step with another
step or an ensemble of
other steps \cite{ep}. Indeed, in the toy model, the entropic repulsion
with the straight step may cause such an interaction, tending
to hinder the growth of the fluctuations and tending to reduce
the value of the characteristic exponent, from 1/4 to 2/11.

Before discussing the range of validity of this regime, it seems useful to
monitor and analyze the fluctuation length $g(t)$ in more
detail. Actually, $g$ is found to be related to
the standard deviation of the histograms $p(\Delta h)(t)$, describing
the probability that the step position differs, after time $t+t_0$, from
its equilibrium value, at $t_0$, by $\Delta h$. More concretely,

\begin{equation}
\Delta h= \sum_{i} (h(i,t+t_0)-h(i,t_0))/L_a
\end{equation}

\noindent
averaging over all active step sites $i$. Typical results are depicted
in Fig. 8 at $k_BT=1.0$ and $L=201$, for different values of $d$ as well as
for isolated steps. The simulation data are compared to the
Gaussian probability distribution with the standard deviation $\sigma$ set 
equal to the fluctuation length $g(t)$. As seen from the figure, there
is a very good agreement, with a very slight asymmetry in the 
simulation histogram for small $d$, as expected due to the
presence of the confining step. It is worthwhile
mentioning that similar observations hold in general, including
periodic boundary conditions as well.

The criterion for encountering the unconventional effective
exponent, $x_{eff} \approx 2/11$, may be argued to be that, for
sufficiently long steps, the
fluctuation length $g_{iso}$ of the corresponding isolated step
is just somewhat smaller than $d$, so that the entropic repulsion
is strong enough to change the unconstrained fluctuations of 
isolated steps to those reflecting the presence of the second
step. If the fluctuation length is too small, or if $d$ is
too large, then one is in the regime of the fluctuations of
isolated steps, (ii). In the opposite case, regime (i), remainders
of the logarithmic growth are seen. Following this consideration, the extent
of regime (iii) depends, especially, on temperature.

A few typical examples are: At $k_BT/J= 1.0$ and $L$ being about
100 to 200, $g_{iso}$ is, for $t$ ranging
from 5000 to 20000 MCS/S, in between 2.5 and
2.7, increasing with $t$. At $k_BT/J= 3.0$, 
$g_{iso}$ rises, in the same time window, from about 6.3 to 
about 7.5. Accordingly, the regime (iii), where $x_{eff}$ is, in a
rather long time interval, near 2/11, is expected to shift to larger distances
$d$ at the higher temperature. We checked the suggestion by varying $d=$ from
5 to 15, at $k_BT/J= 3.0$. Now, the unconventional exponent, 2/11,
is, indeed, found to be quite pronounced at $d$ around
10. Thence, analysis of the correlation length
for isolated steps will lead to reasonable choices of separation
distances to encounter the unconventional growth in the
fluctuations of confined steps.

As has been argued above, the power--law behavior of $G(t)$ with
the exponent 2/11 is not
expected to occur asymptotically, $t,L \longrightarrow \infty$,
in the toy model for finite distance $d$, because $G(t)$ will
be limited by $d$, in contrast to the situation for
isolated steps. The effective exponent, after having displayed
a plateau--like behavior near 2/11, continues to decrease
monotonically, until it finally reaches
zero. Actually, we monitored, at fixed $d$, the time needed
to reach the, supposedly, characteristic
value of $x_{eff}=2/11$, i.e., $t_{2/11}$, as a function
of step length $L$. As had been 
mentioned above, the corresponding time $t_{1/4}$ for isolated
steps is found to diverge as $L$ is going to infinity, showing
that $G(t)$ obeys, in the thermodynamic limit, a simple power--law
asymptotically in time. Now, for confined steps, the simple
power--law describes only a \textit {transient} behavior. As shown in
Fig. 9, for $d=4$ and 5, we obtain, to a good degree of
accuracy, for sufficiently long steps 
$t_{2/11}(L_a)= t_{\infty} -c/L_a$, with
a finite $t_{\infty}$, reflecting the saturation of $G(t)$ at
finite time due to the presence of the second step in the toy model.

\begin{figure}
\vspace{0.8cm}
\resizebox{0.90\columnwidth}{!}{%
  \includegraphics{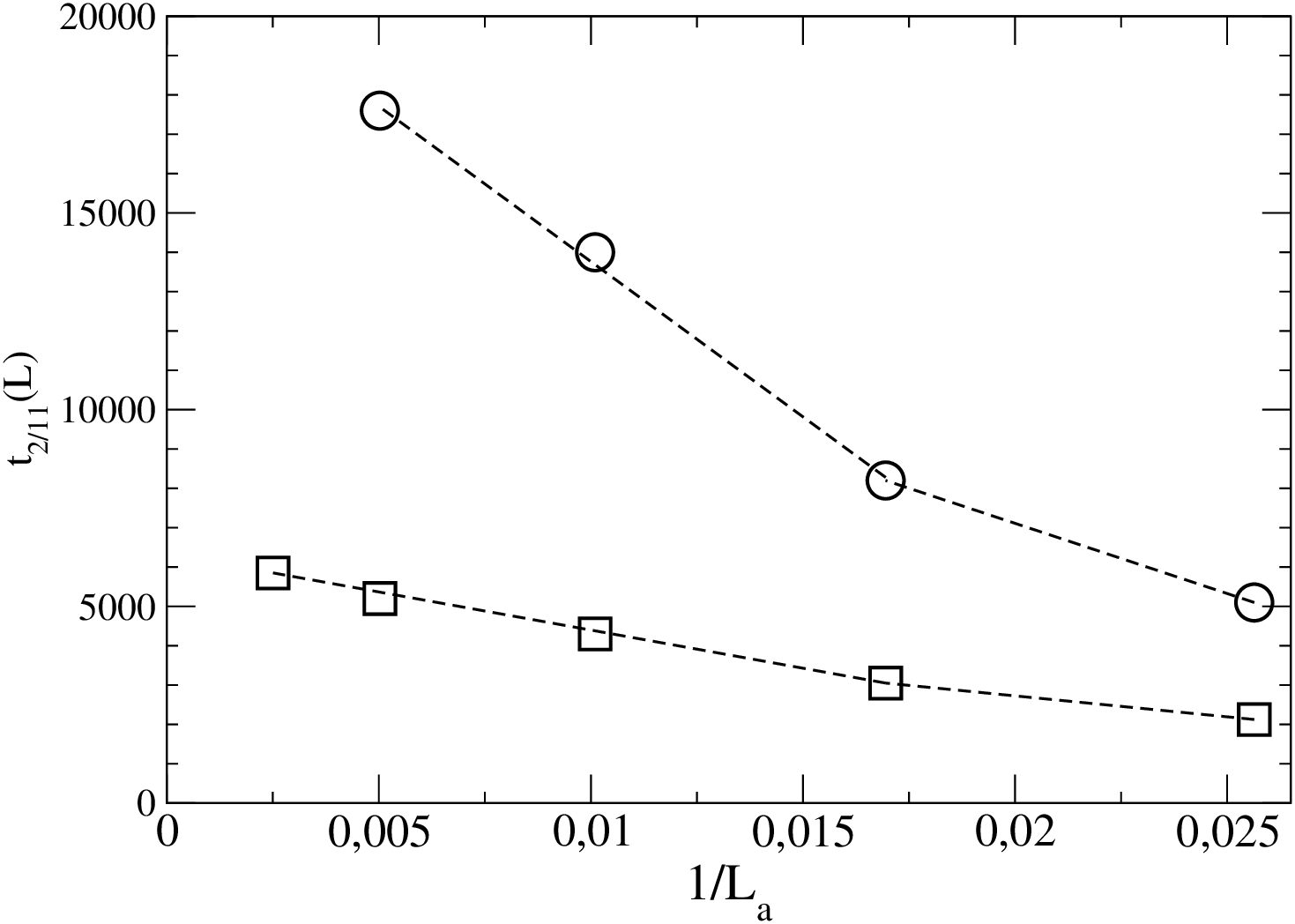}
}
\caption{Finite--size dependence of characteristic time $t_{2/11}$ for
 confined steps with PD for $d=4$ (square) and $d=5$ (circle)
 at $k_BT/J= 1.0$.}
\label{fig:9}
\end{figure}

To observe, possibly, the characteristic value of 2/11 in the
asymptotic limit, the second step may be placed at indefinite distance
from the fluctuating step, with true, physical long--range interactions
between the two steps. Limits had to be taken in
the appropriate order. Note that at larger distances, $d$, longer
and longer runs 
would be needed to determine the limiting
size--dependence, and we refrained from attempting to do it.

For EC, the positions of the fluctuating step
tend to be repelled due the entropic repulsion exerted by
the straight step. Thence, in thermal equilibrium, for pinned
boundary conditions, $h_{eq}(i)$ takes a curved form, somewhat
similar to the one shown in Fig. 1. The maximal height of the
equilibrium shape, at given temperature and step
length, increases
with decreasing $d$. For instance, for steps of
length $L_a$= 199 at $k_BT/J= 3.0$, as depicted in Fig. 7, the
maximal height increases
from about 17 to about 33, when lowering $d$ from 20 to
3. Then, already for small
distances $d$, as depicted in Fig. 7 for $d= 3$, the 
effective exponent $x_{eff}$ does not fall off quickly, 
unlike in regime (i) for PD, with $G(t)$ growing faster
than logarithmically in the time span we studied. As shown in Fig. 7 as
well, for moderate $d$, $d= 20$, there seems to exist a time range in
which $G(t)$ displays a
power--law like behavior with the effective exponent being
close to 2/5, as had been predicted for the border ledge of
a crystal facet \cite{fps,deg1}.

To avoid complications resulting from the curved equilibrium shape
of the step positions, for EC, one may modify the toy
model by introducing another straight
step, placed, symmetrically, at distance $-d$ from the fluctuating
step. For this variant, $x_{eff}$ exhibits similar features as
those shown in Fig. 7, adjusting
the separation distance between the fluctuating and
straight steps. In any event, the EC
case with confinement may deserve further attention. Perhaps, analytical
calculations on the toy models would be
feasible. They would be most welcome.

\section{Summary}

We studied thermal fluctuations, $G(t)$, of isolated and confined steps of
monatomic height on low--index crystal surfaces in the framework
of one--dimensional SOS models. Periphery diffusion, PD, and 
evaporation--condensation, EC, have been analyzed, using Monte
Carlo techniques.

For isolated steps, we found that $G(t)$ grows asymptotically, i.e. for
$t \longrightarrow \infty$, in the limit of indefinitely many step
sites, as $G(t)=at^x$. In the case of PD, the characteristic
exponent $x$ is observed to be 1/4, independent of slope and
curvature as well as temperature of the steps. The prefactor
$a$, on the other hand, varies with the geometry and temperature, being
determined by the energy of the step. The temperature dependence
for initially flat steps agrees well with the one predicted by
Langevin theory. In the case of EC, the exponent $x$ is found 
to be 1/2, independent of slope and temperature.

For confined steps, we considered, following previous simulations, 
a toy model with an
initially flat, fluctuating step in the presence of a second 
fixed, straight step, at initial distance $d$. We mainly studied
PD. For rather large steps, we find, depending on $d$, three
different regimes, where $G(t)$ shows, for quite extended 
time intervals, either a power--law like behavior or 
logarithmic growth. 

The power--law like behavior may be due to unconstrained fluctuations
as for isolated steps, with $x_{eff}$ close to 1/4 or due to 
fluctuations hindered by the entropic repulsion of the confining
step, with $x_{eff}$ being near 2/11. In all cases, fluctuations
are found to satisfy a Gaussian distribution. Comparison of the
standard deviation (or fluctuation length) of the isolated steps
to the separation
distance provides the clue to determine the range of validity
of the various regimes with logarithmic or power--law like
growth of $G(t)$. Eventually, $G(t)$ will be limited
by $d$, so that asymptotically, $t,L \longrightarrow \infty$, the
effective exponent $x_{eff}$ will approach zero, for all finite values of $d$.

In case of EC, the
effective exponent, due to entropic repulsion, may be expected to 
be 2/5, as predicted by related scaling arguments
and Langevin theory. Indeed, a related power--law like growth of
the step fluctuations seems to occur for suitable choices
of distance $d$, temperature, and step length. However, the
simulations of the toy model are
strongly affected by the fact that the average step
position is not conserved for EC. Further studies are
desired for clarification.

\acknowledgments

I should like to thank Theodore L. Einstein for very useful correspondence.

\end{document}